# Observed network dynamics from altering the balance between excitatory and inhibitory neurons in cultured networks

Xin Chen and Rhonda Dzakpasu

Department of Physics, Georgetown University

Department of Pharmacology, Georgetown University Medical Center

**Abstract**

Complexity in the temporal organization of neural systems may be a reflection of the diversity of their neural constituents. These constituents, excitatory and inhibitory neurons, comprise a well-defined ratio *in vivo* and form the substrate for rhythmic oscillatory activity. To begin to elucidate the dynamical implications that underlie this balance, we construct novel neural circuits not ordinarily found in nature and study the resulting temporal patterns. We culture several networks of neurons composed of varying fractions of excitatory and inhibitory cells and use a multi-electrode array to study their temporal dynamics as this balance is modulated. We use the electrode burst as the temporal imprimatur to signify the presence of network activity. Burst durations, inter-burst intervals, and the number of spikes participating within a burst are used to illustrate the vivid differences in the temporal organization between the various cultured networks. When the network consists largely of excitatory neurons, no network temporal structure is apparent. However, the addition of inhibitory neurons evokes a temporal order. Calculation of the temporal autocorrelation shows that when the number of inhibitory neurons is a major fraction of the network, a striking network pattern materializes when none was previously present.

**I. Introduction**

Pattern formation is ubiquitous in biological systems and these patterns are often similar to those found in non-living systems [1-3]. This has piqued the interest of physicists leading them to investigate the relationship between the spatial and temporal patterns and the biological constituents that generate them. Significant inroads have been made in understanding the formation of bacterial colonies as well as characterizing spontaneous and evoked activity in neural circuits [4-6]. Complex activity patterns have been shown to emerge from the self-organization of neurobiological networks and can persist for hours [7,8]. Results from these studies demonstrate that when biological systems interact with nature, intricate and unexpected patterns can form.

In general, complex, i.e., non-periodic, patterns form in open systems that are driven out of equilibrium due to competition over an existing resource [9]. In the brain, there are two types of neurons: excitatory and inhibitory, and it has been suggested that competition for excitatory and inhibitory inputs received by a neuron is essential for healthy brain activity [10]. The brain must operate within a range of activity for which external perturbations do not drive it into the pathological state and it is the balance between excitation and inhibition that maintains this dynamical state [11,12]. This balance is achieved in the cortex by an approximate ratio of 70% excitatory and 30% inhibitory neurons. This ratio appears to be present with minimal variation across a large diversity of species such as rodents, felines and humans [13-17]. Why does this ratio persist and how does it influence pattern formation in neural circuits?

Temporal patterns produced from mixed excitatory and inhibitory networks have been investigated using computational models [18-22]. Anderson et al. used a network of single-compartment excitatory and inhibitory neurons, and by varying the total level of excitation or inhibition; they produced a wide range of dynamics from tonic firing to synchronized bursting [23]. Network connectivity was sparse and the connections were determined randomly. Other groups have fixed the ratio of excitatory/inhibitory cells and have focused on varying the excitatory/inhibitory synaptic connectivity strength as well as the external inputs to each neuron. For example, by using a three-layered network of spiking neurons that represents external input, subcortex and cortex, Xing and Gerstein showed that modulating inhibition had a larger effect on neural receptive fields than excitation [22]. Brunel described analytically the wide range of synchronous as well as asynchronous states that arise from two classes of neural networks: one that has identical characteristics for the excitatory and inhibitory neurons and the other class in which physiological data has been incorporated into the model to differentiate between the two types of neurons [19]. Lastly, Vogels and Abbott used a network of integrate and fire neurons to investigate how signals can turn on in the case when excitatory and inhibitory pathways are imbalanced between the sender and receiver areas of the network [24]. They produce an imbalance by differentially modulating the excitatory and inhibitory pathways between these two regions. However, to the best of our knowledge, no one so far has taken an experimental approach.

This paper describes the intriguing temporal patterns formed in an *in vitro* experiment using networks of cultured neurons. Neural cultures are a simple, reduced two-dimensional system. They may provide insights into basic dynamical network interactions not currently achievable in complex *in vivo* brain preparations. While *in vivo* measurements are clearly the more direct approach to studying physiological dynamics, it is difficult to visualize individual neurons and record single unit electrical activity from *in vivo* three-dimensional networks of neurons. In addition, two-dimensional *in vitro* networks are easily amenable to pharmacological, electrical and genetic modification and they retain many of the properties of *in vivo* networks, such as rich connectivity and complex patterns of activity.

In this study we ask two questions: i) What are some of the complex dynamical network patterns formed by a mixed excitatory and inhibitory culture and ii) how do these patterns change as the balance between excitatory and inhibitory neurons is modulated? Exploiting the ease of manipulation within cultured networks, we start with a single-cell suspension of hippocampal neurons, which are approximately 80% excitatory [25]. To this suspension, we titrate individual neurons from the striatum, which are nearly 100% inhibitory [26,27]. We culture these networks onto an array of electrodes and after a few days, excitatory and inhibitory connections spontaneously form. Increasing the inhibitory fraction in the network increases the heterogeneity of the system. As a result, we suggest that there are more ways for the network to spatially arrange itself. We show that increasing the inhibitory fraction leads to a remarkable temporal correlation pattern.

## II. Methods
### A. Cell cultures

All experimental procedures were carried out in accordance with the Georgetown University Animal Care and Use Committee (GUACUC). Hippocampal and striatal tissue were extracted from embryonic day 18 Sprague-Dawley rats using a protocol modified from ref. 28. Briefly, the neural tissue was finely chopped and digested with 0.1% trypsin followed by mechanical trituration. Upon reaching a single cell suspension, three different concentrations of striatal cells were added to single cell suspensions of hippocampal cells resulting in the following five different cultured networks with an approximate density of 1000 cells/mm$^2$: 100% hippocampus, 80% hippocampus-20% stratum, 67% hippocampus-33% striatum, 55% hippocampus-45% striatum, 100% striatum. Each culture was plated onto a multi-electrode array (Multi-channel Systems, Reutlingen, Germany) that was previously treated with poly-d-lysine and laminin (Sigma, St. Louis, MO). Cultures were maintained in Neuralbasal A medium with B27 (Invitrogen, Carlsbad, CA) with bi-weekly changes and kept in a humidified 5% $CO_2$ and 95% $O_2$ incubator at 37$^o$C.

### B. Electrophysiological recordings

The multi-electrode array (MEA) is composed of 59 titanium nitride electrodes, one reference electrode and four auxiliary analog channels each of which is 30 µm in diameter, arranged on an 8x8 square array. The inter-electrode spacing is 200 µm. Upon plating, the cells in suspension adhere to the silicon nitride substrate of the MEA and after three days electrical activity becomes apparent. We use the MEA1060 preamplifier and sample electrical activity at a 25kHz acquisition rate to allow the detection of multi-unit spikes. The data was digitized and stored on a Dell personal computer (Round Rock, TX). Possible exposure to contaminants was significantly reduced during the experiments by the use of an MEA cover made of a hydrophobic membrane [29]. This membrane provides a tight seal and is permeable to $CO_2$ and $O_2$ and largely impermeable to water vapor. Experiments from at least 10 different networks plated onto MEAs were performed on a heated microscope stage at 37$^o$C for 20 minutes at 18 days *in vitro*, a time point during development in which the network displayed vigorous electrical activity and for which network connectivity is established [30-32].

### C. Data analysis
### i. Spike detection

We remove low frequency components by high-pass filtering all traces at 25 Hz. Extracellularly recorded spikes were detected using a threshold algorithm from Offline Sorter (Plexon Inc., Dallas TX). The threshold is calculated as a multiple of the standard deviation (3.5σ) of the biological noise. No attempt was made to discriminate and sort spikes by electrode because the shape of a spike changes significantly during a burst due to changes in membrane excitability. In addition, for this study we concentrate on network activity and the signal from each electrode suitably reflects these dynamics.

### ii. Burst parameters

We have written proprietary software using Matlab (Mathworks, Natick, MA) to calculate all of our network data analyses. We chose a common temporal feature found in cultured networks in order to study how network dynamics change as the fraction of inhibitory cells is altered. This dominant temporal motif in cultured neural networks is the burst and it represents a collective neural response. In our experiments, we analyze bursts from each individual electrode. After the spike detection process described above each electrode has a resulting spike train, $\tau_{st}(t)$, expressed as:

$$\tau_{st}(t) = \sum_{n=1}^{N} \delta(t - t_n)$$

where N is defined to be the total number of spikes, $t_n$ is the time of the *n*th spike and $\delta(t)$ is a delta function that indicates a spike taking place at time $t = t_n$. The inter-spike interval between spike *n* and spike *n*-1 (n >1) is:

$$\tau^{ISI}_n = t_n - t_{n-1}$$

We define a burst from each electrode to consist of no less than six spikes with a maximum inter-spike interval of 60 ms. As will be described below in Fig. 3 and in the associated text, bursts from networks with greater than a 20% inhibitory cell contribution terminated with a tail of small clusters of (<6) spikes. Setting the "spike-count floor" to be six ensures that these clusters are not erroneously counted as independent episodes. Additionally, defining the maximum inter-spike interval of 60 ms allows us to include the spike clusters as part of the complete episode without including the next episode. This burst identification process results in an M x N matrix where M corresponds to the electrode number and the N's are the time stamps of the bursts.

Using this burst criterion we calculated burst durations for the different cultured networks. In addition, to obtain a finer differentiation between cultures we also calculated the fraction of bursts that have durations less than 100 ms as our burst duration histograms displayed a significant drop off at this value for the 67% H-33% S networks (see below). Next, we calculated the number of spikes per burst for each of the different cultures. We also calculated the fraction of bursts containing 40 spikes or less as the spikes/bursts histograms displayed a significant drop off at this value for the 67% H-33% S networks (data not shown). In addition, there were a considerable number of spikes that were not included in these calculations because they were not part of any burst. Despite the fact that the focus of this study is to measure how network activity is modulated, it has been speculated that whether or not a spike contributes to a burst may be indicative of the information processing efficiency of the network [33-35]. Therefore, we calculated the fraction of spikes that did not participate within a burst for each network. Lastly, to investigate changes in overall network rhythmicity, we calculated the inter-burst intervals and the temporal autocorrelation. For the temporal autocorrelation, we were solely interested in correlations between and not within bursts. As a result, to eliminate contributions to the autocorrelation from intra-burst dynamics, we constructed model bursts to represent each burst that occurred within an electrode. To do this, we define the burst train, $\tau_{bt}(t)$, with a total number of M bursts as,

$$\tau_{bt}(t) = \sum_{m=1}^{M} \prod(\frac{t-t_m}{d_m})$$

where $t_m$ specifies the starting time of the *m*th burst in the $\tau_{bt}(t)$, $\prod(t/d)$ is a rectangular function that specifies the onset of a burst at time $t = t_m$ lasting for duration $d_m$. Using this modified time series, the temporal autocorrelation $\Gamma$ was calculated for each electrode within each network.

## III. Results

All but one of the mixed hippocampal/striatal cultures exhibited robust bursting activity at 18 days *in vitro*. Recordings from all four of the 100% striatal MEAs never displayed electrical activity. These networks are composed solely of inhibitory neurons and we believe the lack of activity is due to the absence of excitatory input. Therefore, results presented will be for the following cultured networks: (i) 100% hippocampus, ii) 80% hippocampus-20% striatum, iii) 67% hippocampus-33% striatum and iv) 55% hippocampus-45% striatum. Lastly, the addition of striatal cells is solely to vary the inhibitory neuron fraction and we will henceforth report our results based on the estimated E/I cell ratio, assuming that the 100% hippocampal networks consist of 80% excitatory (E) and 20% inhibitory (I) neurons.

In Fig. 1, left panel, a phase contrast image of the MEA plated with 80% E – 20% I neurons is presented. The right panel is a screenshot of spontaneous activity as recorded by the MEA. Each box corresponds to one second of activity from one electrode. Each electrode records activity from the neurons in its vicinity and several of these electrodes reveal bursting dynamics. Note the highly synchronous nature from a majority of the electrodes.

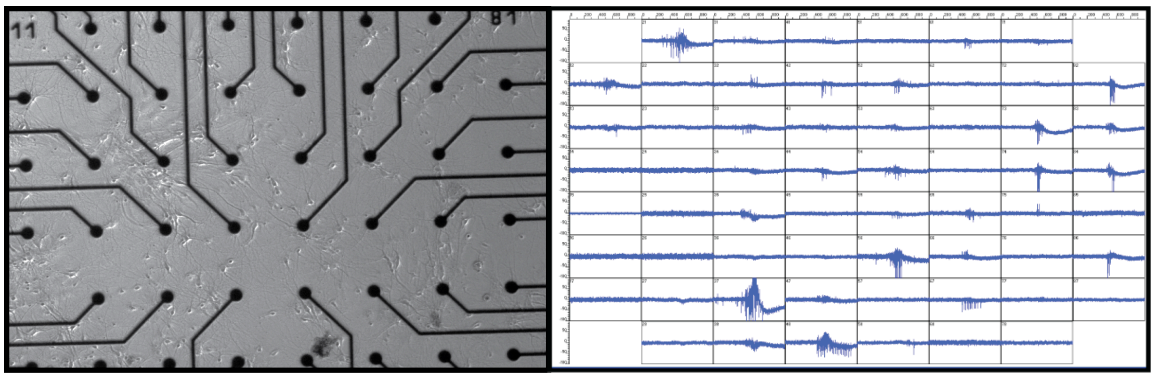

FIG. 1. 80% E-20% I cultures at 18 days *in vitro* on a multielectrode array. Left) Phase contrast image of the cells plated on a multielectrode array (MEA). Right) Screen shot of spontaneous recordings from the MEA. Each box represents one second of recording from one electrode.

### i. Raster plots: an overview of network activity

We created raster plots to highlight how strongly inhibitory cells impact spiking activity on the network level. Fig. 2 presents raster plots of spike trains from 70 seconds of spontaneous activity at 18 days *in vitro* for each cultured network. One row in each panel corresponds to an electrode and in this row each small vertical tick mark is a detected spike. All of the networks display clear evidence of widespread bursting activity albeit with significant differences. The 80% E-20% I network shows a large degree of activity on every electrode throughout the recording. The bursts seem to have similar durations. As the fraction of inhibitory cells increase within the network, the network organizes into regions of longer duration bursts, with increasingly longer quiescent intervals of separation.

On a shorter timescale, the differences in network activity within each culture are more visible. Fig. 3 is a raster plot of 2.5 seconds of spontaneous activity from each cultured network. The bursts for the 80% E-20% I networks continue to look uniform in length and there are many spikes that are not part of a burst. As the fraction of inhibitory cells increases, the bursts begin to broaden in time. However, as we will see below the inter-spike intervals also increase and these "stretched bursts" are no longer one burst, but one long-duration burst with several bursts of short duration that are clustered together: a "super-burst". The constituent spikes of these bursts lose their temporal coherence resulting in the breakup of a burst into multiple "mini bursts".

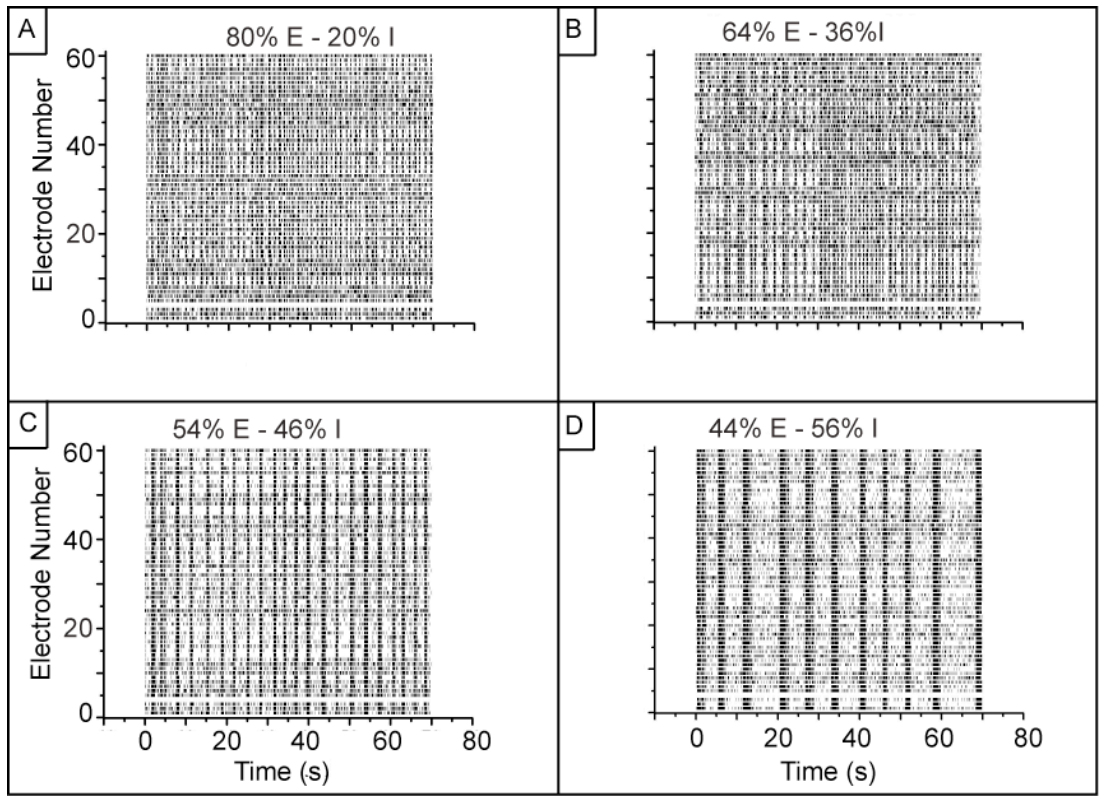

FIG. 2. Raster plots of 70 seconds of spontaneous activity at 18 days *in vitro*. Each panel depicts a different E/I ratio. a) 80% E-20% I. There is a high degree of activity with each electrode displaying bursting activity. b) 64% E-36% I. The activity is very similar to the 80% E-20% I networks. c) 54% E-46% I. The activity is beginning to cluster and organize into large burst structures. d) 44% E-56% I. The activity has changed to bursts of long duration followed by tails of shorter mini-bursts. NOTE: The electrode with no activity is the reference electrode in all cultures and therefore has no signal.

## ii. Burst calculations

We quantified how the bursts were changed as the number of inhibitory cells was increased. The distributions of the burst duration change as the number of inhibitory neurons increases as seen in Fig. 4 and table 1. Initially, the distribution appears to be log-normal (Fig. 4A) and as the number of inhibitory cells increases, the distribution expands and the standard deviation increases by a factor of three (Fig. 4B). The profile for the 64% E-36% I cultures is not dissimilar from the 80% E-20% I network. However, when the fraction of inhibitory cells increases to 46%, the shape of the distribution as well as the standard deviation changes significantly (Fig. 4C). This difference is amplified in the 44% E-56% I network as this distribution appears to be exponential (Fig. 4D). Additionally, in the 44% E network, there is a tail of long durations that is not present in the other networks (Fig. 5). Lastly, despite the fact that there is a 30% increase in the mean burst duration from the 80% E to 44% E cultures, the median burst duration decreases by 40% suggesting that there are more bursts of short duration present in the 44% E neural cultures.

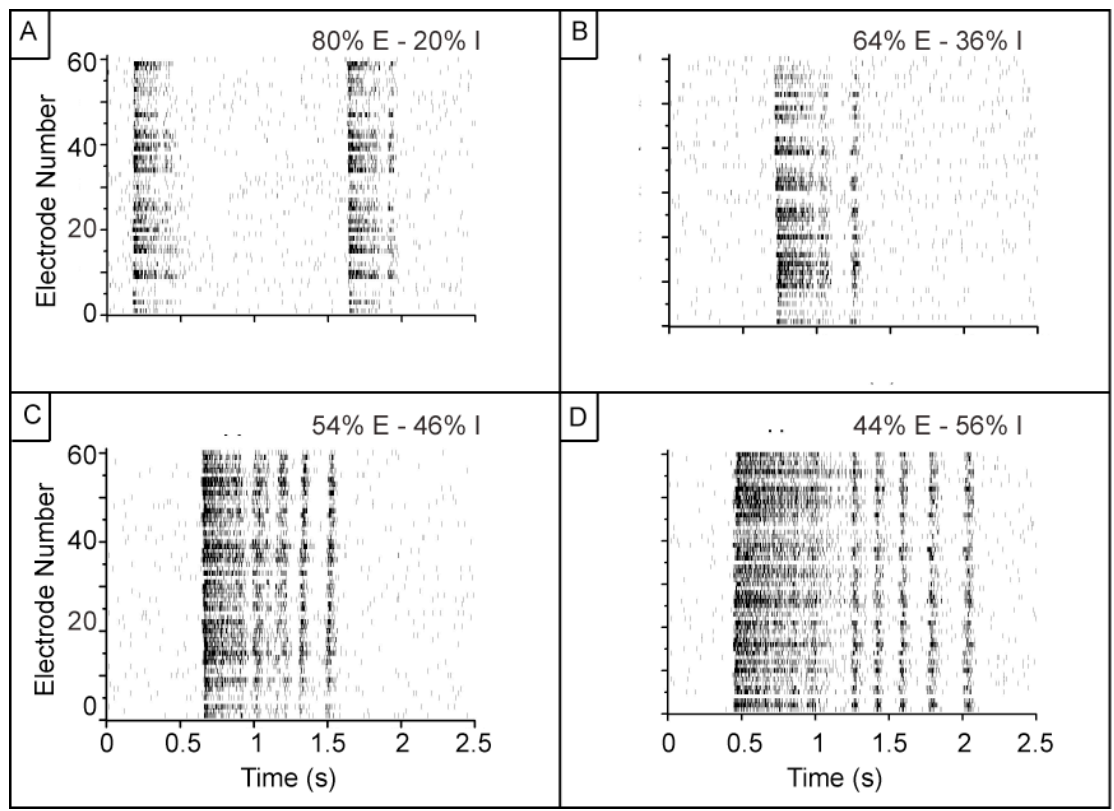

FIG. 3. Expanded time scale of raster plot of spontaneous recordings at 18 days *in vitro*. Each panel refers to a different E/I ratio. a) 80% E-20% I. The bursts are largely uniform in their length. Also, there are many spikes that are not part of the burst. b) 64%E-36%I. The bursts are beginning to lengthen in time. c) 54% E-46% I. The bursts are now quite long and will become several small bursts. d) 44% E-56% I. The burst is breaking up and the long tail of multiple short bursts is now apparent. NOTE: The electrode with no activity is the reference electrode in all cultures and therefore has no signal.

| Network Composition (Number of MEAs) | Mean (ms) | Median (ms) | Standard Deviation |
|---|---|---|---|
| 80% E – 20% I (2) | 170.2 | 168 | 87.7 |
| 64% E – 36% I (3) | 165.4 | 159.9 | 95.5 |
| 54% E – 46% I (3) | 190.4 | 110.2 | 171.0 |
| 44% E – 56% I (4) | 225.6 | 102.3 | 256.3 |

Table 1. Burst Duration statistics

To study this further, we calculated the ratio of bursts that were 100 ms or shorter for each of the networks (Fig. 6). This ratio increases as the number of inhibitory cells increases with a two-fold increase from the 80% E to 44% E cultures.

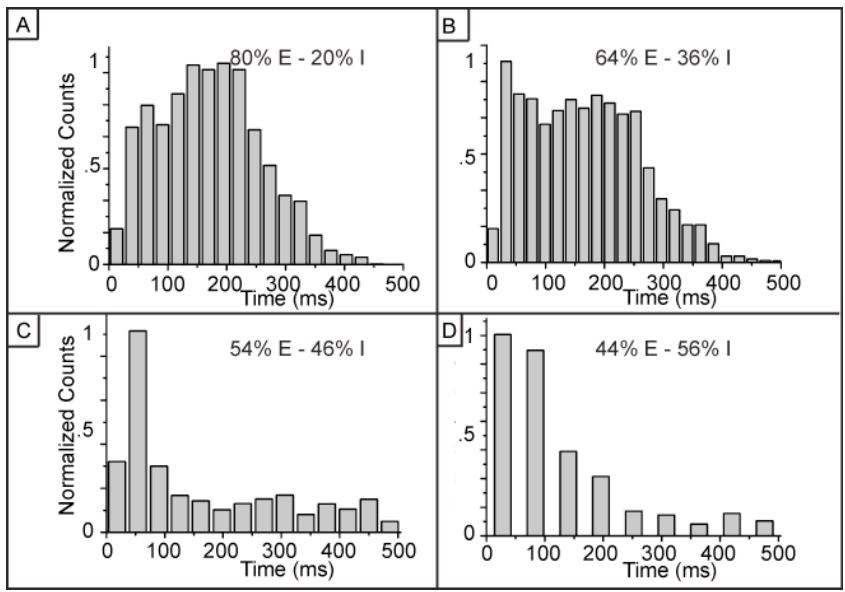

FIG. 4. Normalized burst duration histograms for spontaneous activity for the MEAs of each E/I ratio. a) 80% E-20% I. The distribution appears to be log-normal. b) 64% E-36% I. The distribution is similar to panel A with some broadening. c) 54% E-46% I. There is a transition to an exponential-like distribution with a marked shift towards shorter durations, less than 100 ms. d) 44% E-56% I. Most of the bursts are of short duration.

Next, we calculated the number of spikes per burst for all of the networks (Fig. 7A). For all of the cultures, the inter-quartile distributions are tightly clustered. However while the average number of spikes per burst increased 36% from the 80% E-20% I cultures to the 44% E-56% I cultures the median dropped by 22% and there was a 1.5-fold increase in the standard deviation (table 2). As observed in Fig. 4, the 44% E-56% I cultures displayed an increase in bursts of short durations, suggesting that there may be bursts with low numbers of spikes. Therefore we calculated the fraction of bursts that contain 40 or less spikes (Fig. 7B). There was a 30% increase in the bursts with 40 or less spikes from the 80% E-20% I cultures to the 44% E-56% I cultures.

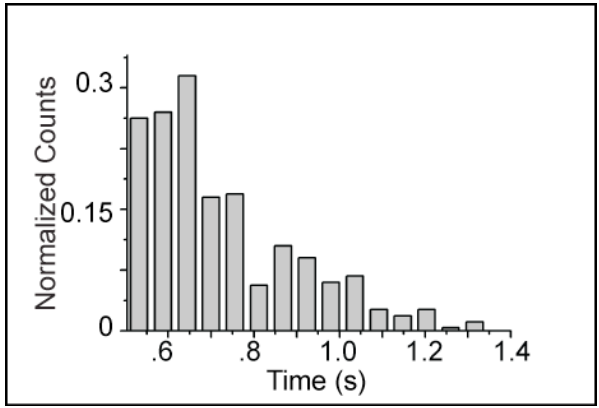

FIG. 5. Normalized long burst duration histogram for 44% E-56% I cultures. Extending the durations in Fig 4D there is a fraction of this cultured network having long burst durations as compared to the 80% E-20% I networks. There are more "super-bursts" with long durations, as the increased number of inhibitory cells appears to break apart the bursts.

We are studying the network response to changes in inhibitory cell populations and therefore only those spikes that participate within a burst are considered in these analyses. However, it is evident from the raster plots of Fig. 3 that there are large numbers of spikes that occur outside of bursts. A proposed role for bursting in neural circuits is that a tight barrage of spikes may be more efficient to propagate information with a diminished role in information transmission for individual spikes [33-35].

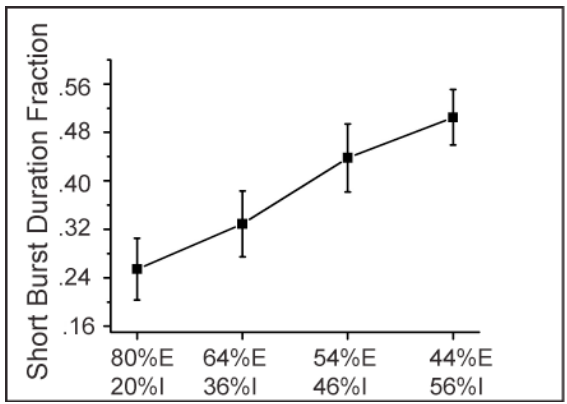

FIG. 6. Short burst duration fraction. The ratio of burst durations less than 100 ms with respect to all burst durations. The fraction of short bursts increases nearly linearly as the number of inhibitory cells increase. There are approximately twice as many short bursts in the 44% E networks than in the 80% E networks.

We calculated the fraction of spikes that do not participate within a burst for the different cultures in Fig. 8. As the inhibitory cell contribution increases, more spikes get recruited in bursts. There is a 50% decrease in the number of spikes that do not participate in a burst for the 44% E-56% I cultures as compared to the 80% E-20% I cultures. Despite the fact that the burst durations are shorter in the 44% E-56% I cultures, more spikes are engaged in bursting activity.

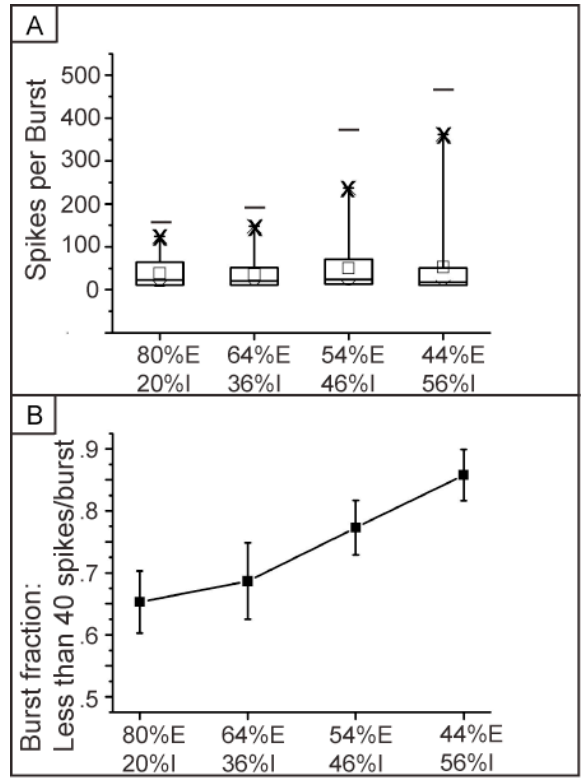

FIG. 7. Number of spikes per burst. A) The boundaries of the box define the inter-quartile region, i.e., $25^{th}$-$75^{th}$ percentile (average spikes/burst for each MEA within each E/I ratio: one-way ANOVA, $p<0.05$). Within each box the line is the median and the square signifies the mean. Outside the box, each dash defines the maximum burst duration and below the dash, the duration marked by the boldface x is the $95^{th}$ percentile point. The width of the distributions increases as the number of inhibitory cells increase. The variability also increases dramatically. There is also an inverse relationship between the mean and the median implying that the 44% E networks have a considerable number of bursts containing small numbers of spikes. B) Bursts with 40 or less spikes. The number of spikes within a burst is greatly reduced when the amount of inhibitory cells is increased.

### iii. Network periodicities

Inter-burst intervals were calculated for each of the different networks (Fig. 9). The majority of the inter-burst intervals for the 80% E-20% I networks are distributed between 0-2 seconds with a mean near 1-1.5 seconds (Fig. 9A). This distribution begins to shift towards shorter inter-burst intervals when the fraction of inhibitory cells increases to 36% (Fig. 9B). Additionally, a second distribution begins to populate intervals from 2-10 seconds. When the concentration of inhibitory neurons reaches 46%, there is a contraction in the distribution of short intervals with virtually all intervals occurring between 0 and 500 ms (Fig. 9C). The long interval region is clustered with intervals ranging from 2-4 seconds. A bimodal distribution is now well established; 50% of the inter-burst intervals are clustered between 0-500 ms and the other half is clustered within 2-4 seconds. This effect is also evident in the raster plots of Figs. 3C and 4C. As the bursts are lengthening in duration, short bursts form and break away from the originating, but now elongated, bursts with short inter-burst intervals. This "super-burst", i.e., the elongated burst and collection of mini-bursts, has the longer inter-burst interval. Increasing the inhibitory contribution to 56% amplifies this effect. One very narrow range of inter-burst intervals, corresponding to the mini-bursts, is less than 1 second, with the majority of inter-burst intervals ranging from 0-500 ms (Fig. 9D). The "super-burst" intervals span a range from 4-10 seconds. Note that in the 80% E-20% I network, only 5% of the inter-burst intervals have intervals longer than 2 seconds (Fig. 9A). As the percentage of inhibitory cells grows to 56% there is a five-fold increase in the number of inter-burst intervals that are greater than two seconds (Fig. 9D). In the latter network there are more "super-bursts" with longer inter-burst intervals.

| Network Composition | Mean | Median | Standard Deviation |
|---|---|---|---|
| 80% E – 20% I | 38.5 | 23 | 33.5 |
| 64% E – 36% I | 36.2 | 21 | 35.0 |
| 54% E – 46% I | 50.1 | 24 | 50.1 |
| 44% E – 56% I | 52.7 | 18 | 77.9 |

Table 2. Spikes per burst statistics

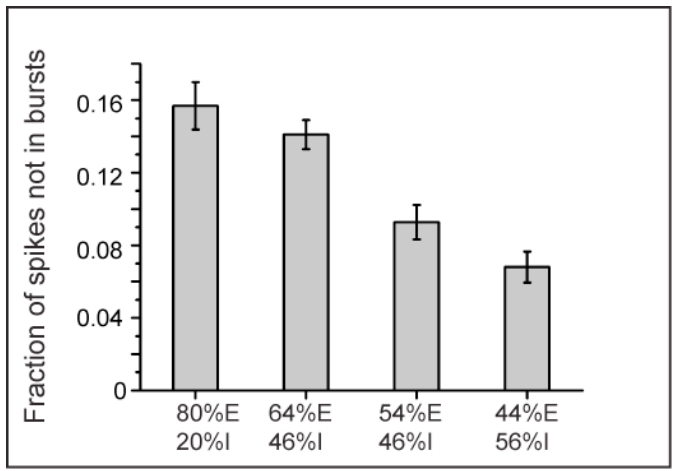

FIG. 8. "Extra-burst" spikes. We calculated the fraction of spikes that do not participate in a burst for each of the E/I cultures. The fraction of non-bursting spikes decreases by a factor of 2 as the number of inhibitory neurons increases. This suggests that inhibitory neurons may improve the efficiency of information transfer by recruiting more spikes in bursts and leaving less "errant" spikes.

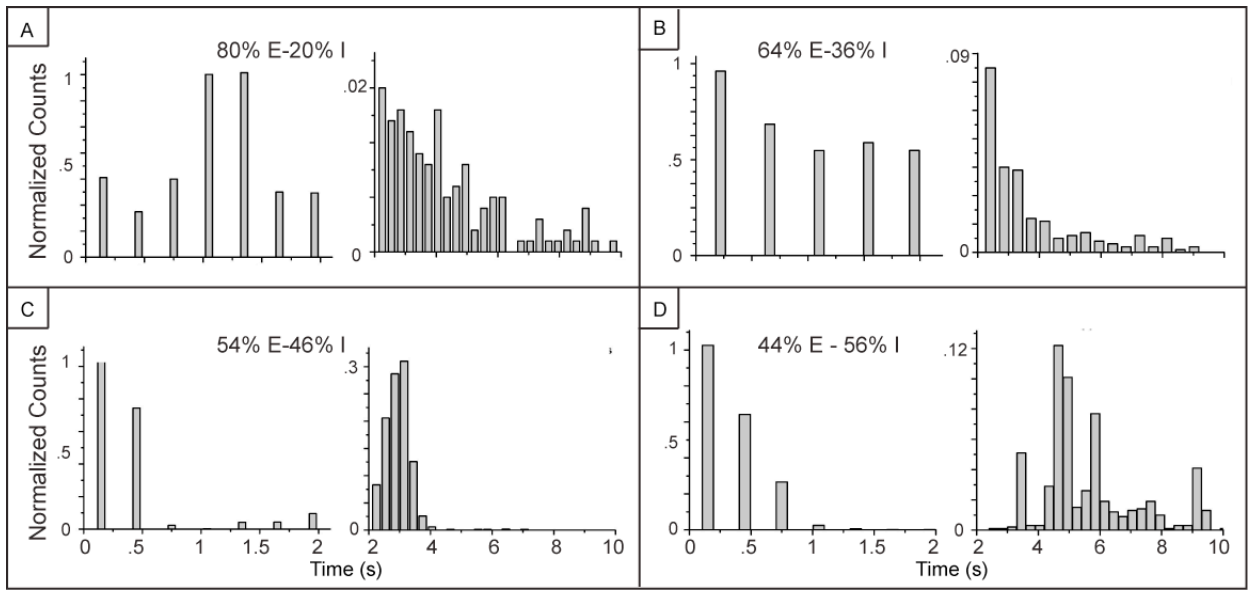

FIG. 9. Normalized inter-burst intervals for each E/I neural ratio. Intervals span two ranges: 0-2 seconds and 2-10 seconds. A) 80% E-20% I. The inter-burst intervals peak between 1-1.5 seconds. Very few intervals are between 2-10 seconds. B) 64% E-36% I. The inter-burst intervals begin to shift towards shorter intervals that are less than one second. There are more intervals in the 2-10 second range. C) 54% E-46% I. There is a marked shift with a cluster of inter-burst intervals between 0-500 ms. A bimodal distribution begins to appear. There is a considerable fraction of intervals in the 2-10 second region. D) 44% E-56% I. This shift is very striking as there are essentially no inter-burst intervals between 1.5-2 seconds. The bimodal distribution is now well established and there is a spread of intervals between 2-10 seconds. These long intervals represent the long "super-bursts" that are seen in the raster plots of Figs. 2 and 3.

Lastly, we calculated the temporal autocorrelation $\Gamma$ for all electrodes within each network to investigate the burst temporal structure within each electrode of the different networks (Fig. 10). When there are relatively few inhibitory neurons in both the 80% E and 64% E networks, no obvious periodicities are present. As the number of inhibitory cells increases, a temporal ordering begins to emerge within the network. The numerous small peaks disappear and a periodic temporal pattern is clearly apparent.

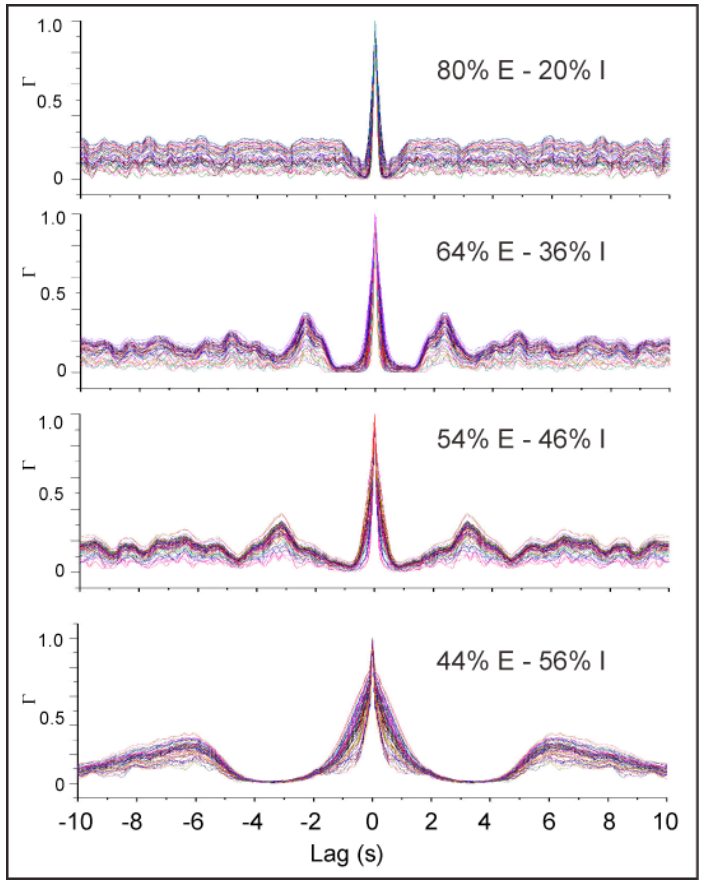

FIG. 10 (color online). Temporal autocorrelation. Normalized autocorrelations were calculated from all active bursting electrodes from each different network. Each graph depicts a different E/I ratio. Within each graph is the temporal autocorrelation of each active electrode. For the 80% E-20% I networks, there are no clear periodicities, however as the number of inhibitory cells increases, a clearly defined temporal structure appears.

## IV. Discussion and Conclusions

We describe spatio-temporal patterns that form from varying the number of inhibitory neurons in an *in vitro* network of cultured neurons. These results are exciting and thought provoking as they clearly demonstrate the profound effect inhibitory cells have on spontaneous network activity. Our analyses suggest that since the presence of inhibitory neurons greatly influences all aspects of burst dynamics, they undoubtedly play an influential role in the overall transmission of information in neural networks. As their numbers increase in the network, the burst durations shorten, however more spikes are recruited into bursting activity. While the role of bursts in neural circuits is still an open question, previously published reports have suggested they may facilitate efficiency of information propagation [33-35]. Based upon these works, we postulate that the "extra-burst" spikes may be thought of as noise in the system and the increase of inhibitory neurons increases the signal-to-noise ratio, i.e., more bursts. In addition, the number of spikes per bursts decreases as the number of inhibitory cells increase.

This suggests that initially, in the 80% E-20% I networks, there is a tightly correlated cluster of neurons whose activity can be considered to be a single functional module. As the number of inhibitory neurons increases, the size of this functional cluster decreases - there are simply fewer spikes within a burst as the inhibitory fraction increases - but this smaller cluster utilizes more of the available spikes. There are fewer "errant" spikes suggesting that the increase in inhibitory neurons may enhance the propagation of information.

The process of culturing networks of neurons results in the random formation of excitatory and inhibitory connections. We start with neuronal solutions that contain different numbers of inhibitory neurons. When we pour each mixture onto the MEA, the inhibitory neurons randomly distribute on the substrate; we do not influence their spatial positions. Therefore, we speculate that the addition of the inhibitory neurons results in more ways for the network to self-organize. While further studies are needed to elucidate the dynamical mechanisms due to the inhibitory neuron influence, our results strongly suggest that their presence has dramatic effects on network temporal patterning. We show in the temporal autocorrelation that when there is excessive excitation, as in the case of the 80% E-20% I networks, no periodic structure is present. However, at a critical inhibitory concentration, periodicities or temporal order appears. A new dynamical pattern emerges when spatial disorder and system heterogeneity increases.

Acknowledgements
The authors are deeply grateful to Jian-Young Wu, Stefano Vicini and Ernest Barreto for their very fruitful and illuminating discussions.

References:

1. E. Ben Jacob, H. Levine, Nature **409**, 985 (2001).
2. J.L.P. Velazquez, J. Biol. Phys. **35**, 209 (2009).
3. S. Arouh, H. Levine, Phys. Rev. E. **62** (1 Pt B), 1444 (2000).
4. J.M. Beggs, D. Plenz, J Neurosci **23**, 11167 (2003).
5. E.D. Gireesh, D. Plenz, Proc Natl Acad Sci **105**, 7576 (2008).
6. R. Quian Quiroga, S. Panzeri, Nat Rev Neurosci **10**, 173 (2009).
7. J.N. MacLean, B.O. Watson, G.B. Aaron and R. Yuste, Neuron **48**, 811 (2005).
8. Y. Ikegaya, G. Aaron, R. Cossart, D. Aronov, I. Lampl, D.Ferster, and R. Yuste, Science **304**, 559 (2004).
9. A. M. Turing, Philos Trans R Soc Lond B **237**, 37 (1952).
10. K. D. Miller, Neuron **17**, 371 (1996).
11. G.G. Turrigiano, S.B. Nelson, Nat Rev Neurosci **5**, 97 (2004).
12. E. Marder, A. A. Prinz, Bioessays **24**, 1145 (2002).
13. A.J. Rockel, R. W. Hiorns and T. P. Powell, Brain **103**, 221 (1980).
14. S. H. Hendry, E. G. Jones, J. DeFelipe, D. Schmechel, C. Brandon and P.C. Emson, Proc Natl Acad Sci USA **81**, 6526 (1984).
15. A. Larkman and A. Mason, J Neurosci **10**, 1407 (1990).
16. J. G. Parnavelas, J. A. Barfield, E. Franke and M.B. Luskin, Cereb Cortex **1**, 463 (1991).
17. D. Contreras, Neural Networks **17**, 633 (2004).
18. N. Brunel, J Physiol Paris **94** 445 (2000).
19. N. Brunel, J Comput Neurosci, **8** 183 (2000).
20. Y. Aviel, D. Horn and M. Abeles, J Neural Comput, **17** 691 (2005).
21. Y. Roudi and P. E. Latham, Plos Comput Biol, **3** 1679 (2007)
22. J. Xing and G. L. Gerstein, J Neurophysiol **75** 184 (1996).
23. W.S. Anderson, P. Kudela, J. Cho, G.K. Bergey and P.J. Franaszczuk, Biol Cybern **97** 173 (2007).
24. T. P. Vogels and L. F. Abbott, Nat Neurosci **12** 483 (2009).
25. M.L. Feldman, *Cellular components of the cerebral cortex* (Plenum Press, New

York, 1984).
26. G. A. Graveland and M. DiFiglia, Brain Res **327** 307 (1985).
27. A.C. Kreitzer and R.C. Malenka, Neuron **60** 543 (2008).
28. D. T. Pak, S. Yang, S., Rudolph-Correia, S., E. Kim and M. Sheng, Neuron, **31**, 289 (2001).
29. S. M. Potter and T. B. DeMarse, J Neurosci Methods **110** 17 (2001).
30. R. Segev, I. Baruchi, E. Hulata and E. Ben-Jacob, Phys Rev Lett **92** 118102 (2004).
31. D. A. Wagenaar, J. Pine and S. M. Potter, BMC Neurosci **7** 11(2006).
32. J. Stegenga, J. Le Feber, E. Marani and W. L. Rutten, IEEE Trans Biomed Eng **55** 1382 (2008).
33. J. E. Lisman, Trends Neurosci **20**, 38 (1997).
34. E. M. Izhikevich, N. S. Desai, E. C. Walcott and F. C. Hoppensteadt, Trends Neurosci **26**, 161 (2003).
35. G. Buzsáki, J. Csicsvari, G. Dragoi, K. Harris, D. Henze and H. Hirase, Cereb Cortex **12**, 893 (2002).